\documentstyle[aps,epsfig]{revtex}

\newcommand{\bold}[1]{{\bf{ #1}}}              
\newcommand{\beq}{\begin{equation}}
\newcommand{\eeq}{\end{equation}}
\newcommand{\beqar}{\begin{eqnarray}}
\newcommand{\eeqar}{\end{eqnarray}}
\newcommand{\bfig}{\begin{figure}}
\newcommand{\efig}{\end{figure}}

\newcommand{\bd}{\begin{itemize}} 
\newcommand{\ed}{\end{itemize}} 
\newcommand{\bc}{\begin{center}}
\newcommand{\ec}{\end{center}}
\newcommand{\be}{\begin{equation}}
\newcommand{\ee}{\end{equation}}

\newcommand{\ba}{\begin{array}}
\newcommand{\ea}{\end{array}}

\begin{document}
\title{Microscopic theories of neutrino-$^{12}C$ reactions}
\author{C. Volpe$^{a}$, N.Auerbach$^{b}$, G.Col\`o$^{c}$,
  T.Suzuki$^{d}$, N. Van Giai$^{a}$}
\address{$^{a)}$
  Groupe de Physique Th\'{e}orique, Institut de Physique Nucl\'{e}aire,
F-91406 Orsay Cedex, France\\
$^{b)}$
School of Physics and Astronomy, Tel Aviv University, Tel Aviv 69978,
Israel \\
$^{c)}$
Dipartimento di Fisica, Universit\`a degli Studi, via Celoria 16, 
20133 Milano, Italia\\
$^{d)}$ 
Departement of Physics, College of Humanities and Science,
Nihon University, Sakurajosui 3-25-40, Setagaya-ku, Tokyo 156, Japan \\
}
\maketitle

\begin{abstract}
In view of the recent experiments on neutrino oscillations performed
by the LSND and KARMEN collaborations as well as of future experiments,
we present new theoretical results of the flux averaged
$^{12}C(\nu_e,e^-)^{12}N$ and
$^{12}C(\nu_{\mu},{\mu}^-)^{12}N$ cross sections.
The approaches used are charge-exchange RPA, charge-exchange
RPA among quasi-particles (QRPA) and the Shell Model.
With a large-scale shell model calculation the exclusive
cross sections are in nice agreement with the experimental values for
both reactions. The inclusive cross section for $\nu_{\mu}$ coming
from the decay-in-flight of $\pi^+$ is $15.2 \times 10^{-40}~cm^2$
(when Hartree-Fock
wavefunctions are used), to be compared to the experimental value of
$12.4 \pm 0.3 \pm 1.8 \times 10^{-40}~cm^2$, while the one due to $\nu_{e}$
coming from the decay-at-rest of $\mu^+$ is $16.4 \times 10^{-42}~cm^2$
which agrees within experimental error bars with the measured values.
The shell model prediction for the decay-in-flight neutrino
cross section
is reduced compared to the RPA one,
namely $19.2 \times 10^{-40}~cm^2$.
This is mainly due to the different kind of correlations
taken into account in the calculation of the spin modes
(in particular, because of the quenching in the $1^+$ channel)
and partially due to the shell-model
configuration basis which is not large enough, as
we show using arguments based on sum-rules.
Results for exclusive and inclusive muon capture rates and beta decay
are given and are close to the experimental findings.

\end{abstract}

\section{Introduction}
For many years, weak processes in nuclei such as
beta decay and muon capture
have been studied on the one hand to
deepen our knowledge
of the weak interaction in nuclei and on the other hand to
yield information on nuclear structure.
Both aspects are important 
when one considers other weak processes in nuclei,
namely
reactions induced by scattered neutrinos.
A description of the latter processes is not only important 
in our attempts to better understand  the nature of such reactions but also
it has significant practical importance in current experimental studies of
neutrinos.
In fact,  nuclei are often used as neutrino detectors 
so that the knowledge of reactions induced by neutrinos on nuclei
becomes a crucial step for the interpretation
of experiments on neutrinos, such as the ones aiming to go beyond the
Standard Model looking for neutrino oscillations and 
masses or those measuring solar
neutrinos to test the Standard Solar Model.

A clear example is given by
the recent experiments performed both by the LSND and the KARMEN
collaborations, looking for
$\nu_{\mu} \rightarrow \nu_{e}$ \cite{LSND1,KR0},
$\bar{\nu}_{\mu} \rightarrow \bar{\nu}_{e}$ \cite{LSND2,KR1}
or ${\nu}_{\mu} \rightarrow {\nu}_{x}$ \cite{KR2}
oscillations
with
neutrinos produced by accelerators.
The detectors used in these measurements are mainly composed
of protons and $^{12}C$.
Reactions of neutrinos on this nucleus are used to check neutrino
fluxes and efficiencies \cite{LSNDvmu}.
In the  ${\nu}_{e} \rightarrow {\nu}_{x}$ disappearance
experiment of Ref.\cite{KR2}
with ${\nu}_{e}$ coming from the Decay-At-Rest (DAR)
of $\mu ^{+}$,  the charged-current (CC)
reaction $\nu_{e} + ^{12}C \rightarrow e^{-} + ^{12}N_{g.s.}$
is used to detect neutrinos. 
In Ref.\cite{LSND1} the $\nu_{\mu} \rightarrow \nu_{e}$ appearance
experiment 
with $\nu_{\mu}$ coming from the Decay-In-Flight
(DIF) of $\pi^+$, $^{12}C$ is used to detect neutrinos
via the CC reaction
$\nu_{e} + ^{12}C \rightarrow e^{-} + ^{12}N$.
In both cases,
the extracted oscillation probabilities rely directly on
the knowledge of the cross section for these reactions.
The exclusive cross section for 
$\nu_{e} + ^{12}C \rightarrow e^{-} + ^{12}N_{g.s.}$
(where $^{12}N$ is left in the ground state)
has been measured by different collaborations
\cite{LSNDve,Allen,KAR_ve_Ngs} and its value can be obtained
in a model independent way
by using form factors deduced from the measurements of
related weak processes
such as beta decay and muon capture as it is done in
the Elementary Particle Theory (EPT) \cite{KuboEPT}.
Concerning the inclusive cross section for
$\nu_{e} + ^{12}C \rightarrow e^{-} + ^{12}N$
(where $^{12}N$ is left either in the ground state or in an excited state),
because of the universality of the weak interaction we expect this
reaction to be described by the same effective hamiltonian as the reaction
$\nu_{\mu} + ^{12}C \rightarrow \mu^{-} + ^{12}N$
with $\nu_{\mu}$ coming from the DIF of $\pi^{+}$.
A problem concerning this reaction
has emerged which has
been extensively investigated recently,
namely the theoretical cross section overestimates
the experimental value \cite{LSNDvmu} by about $50 \%$
within charge-exchange RPA \cite{KL95,Vorov} or by about $ 30-40 \%$ 
in \cite{Singh}. 
An attempt has also been made within an extension of the
EPT \cite{Mintz}, but it is based on several assumptions which have not been
tested yet \cite{KL97}.
In a recent shell model calculation \cite{Hayes} 
the value of the cross section is reduced in apparent agreement
with the experimental value.
The reaction cross section with neutrinos coming from the
DIF of $\pi^{+}$ has
been measured only by the LSND collaboration up to now \cite{LSNDvmu,Imlay}.
Because 
the same detector used for these measurements has been used
to measure the neutrino oscillations,
it is important to know
what we can say about these reaction cross
section
from the theoretical point of view, keeping in mind that every nuclear
structure model necessarily contains approximations.

The weak interaction in nuclei being well known,
an accurate prediction of these cross sections is
a challenging problem from the nuclear structure point of view.
In fact,
these observables
rely on transition densities which can be obtained only in
calculations which
take into account the different aspects of the structure of the
nuclei involved  in these reactions.
First, it is known for a very long time \cite{CK65} that
$^{12}C$ is not a true closed sub-shell nucleus or, in other words, that
the ground state wave function can be only described by an intermediate
coupling scheme and contains configuration mixing, including
deformed components.
Second, if on one hand in the DAR experiments the neutrinos have impinging energies of
the order of several tens of $MeV$, in the DIF experiment
the energy goes up to
around $300~MeV$. As a consequence, the DAR reaction cross section
is dominated by the Gamow-Teller (GT) transition to the ground state of $^{12}N$ for which
information on the transition probabilities can be obtained from other related
weak processes like $^{12}N ({\beta}^{+}) ^{12}C$,
$^{12}B ({\beta}^{-}) ^{12}C$ and
$^{12}C ({\mu}^{-}) ^{12}B$.
On the contrary, in the DIF reaction 
the energy and momentum transferred to the nucleus is quite large so that
$^{12}N$ can be left in an excited state of several tens of
MeV, that is, in and above the giant resonance region.
If the transition to $^{12}N_{gs}$ represents
2/3 of the total reaction cross section in the DAR case,
the cross section given by
transitions to the excited states is 200 times larger than
the transition
to the $^{12}N_{gs}$, if the neutrinos come from the DIF of pions.
This means that to have an accurate prediction for these cross sections
we need nuclear structure models which are not only capable  of
taking into account configuration mixing in the ground
state wave function of $^{12}C$ but also
of describing high-lying states in $^{12}N$.
This is a quite difficult task within the present nuclear structure
models.

The microscopic theoretical approaches used so far are either the 
charge-exchange Random-Phase-Approximation (we will call it from now on
simply RPA)
\cite{pnRPA} or the shell model \cite{Heyde}. The former 
includes only partially
configuration mixing in the ground state 
while it can easily
include  
high-lying one particle-one hole (1p-1h)
configurations. On the other hand the shell model can give a
good description of the ground state wave function 
whereas the prediction of high-lying states requires a 
large model-space which may be very difficult to treat numerically.

In this paper we try to improve the existing calculations within the two
microscopic approaches just mentioned. First, we use 
a charge-exchange RPA approach applied to quasi-particles
(we will call it from now on simply QRPA), in order to improve the poor
description of the ground state of $^{12}C$ within RPA. 
The configuration mixing is introduced by including ``{\it ad hoc}''
pairing correlations in this nucleus.
This is done to see if the configuration mixing that is missing 
in the RPA ground state wave function
can be at the origin of the discrepancy between
the measured reaction cross section and the theoretical 
 predictions
as it was first suggested in \cite{Vorov}.
It was shown in \cite{Vorov} that the inclusion of fractional occupancies
within the RPA approach reduces
the exclusive cross section 
 from a factor 3-4 to $50 \%$ discrepancy.
In Ref.\cite{KL99} an RPA calculation with partial occupancies
was performed and it led to 
similar results for the exclusive cross section 
while the inclusive cross
section  was shown to decrease by
a few percent. However, 
the authors of \cite{Vorov,KL99} did not show the results of a full
QRPA calculation as we do here. 

Next, 
we perform shell model calculations
within a large model space. Our space is 
larger than that actually used in \cite{Hayes} where
the results in the same space we use are obtained by extrapolation
only.
We will compare our results to the experimental findings 
\cite{LSNDvmu,LSNDve,Allen,KAR_ve_Ngs,Imlay} 
and to the other microscopic
 theoretical predictions \cite{Vorov,Hayes,KL99}.
We will conclude by summarizing the present status of the problem.

In section II we briefly review the general theory describing
neutrino scattering on nuclei, 
muon capture and beta decay.
In section III we present the essential features of the microscopic models
used in our calculations, i.e. the QRPA and the shell model.
In section IV results from the above mentioned weak processes
are presented focusing in particular on the cross sections
for the reactions $\nu_{l} + ^{12}C \rightarrow l + ^{12}N$ ($l=e,~\mu$),
both inclusive and exclusive, with $\nu_e$ coming from DAR of
$\mu^{+}$ and with $\nu_{\mu}$ coming from the DIF of
$\pi^{+}$. 
Conclusions are drawn in section V.

\section{General Theory}

\noindent
The theoretical framework to
study nuclear responses to weak probes is discussed extensively
in the literature.
A detailed description can
be found in \cite{Uber72}.
Here we will just present the main ingredients of the
calculation necessary for the discussion. 

The general expression for the cross section of the reaction 
$\nu_{l} + ^{12}C \rightarrow l + ^{12}N$ ($l=e,~\mu$) is \cite{Kubo}
\begin{equation}
\sigma=(2\pi)^{4}\sum_{f}\int d^{3}p_{l}\delta(E_l+E_f-E_{\nu}-E_i)
\vert  \langle l(p_l);f \vert {\it H_{eff}} \vert
\nu_l(p_{\nu});i \rangle \vert ^{2},
\label{e:1}
\end{equation}
where $E_{f}$ ($E_{i}$) is the energy of the final (initial) nuclear
state,
$E_{\nu}$ ($p_{\nu}$)
is the incident neutrino energy (momentum) and
$E_l$ ($p_{l}$)
is the outgoing lepton
energy (momentum).
 The effective single-particle hamiltonian ${\it H_{eff}}$ is derived by carrying
out the Foldy-Wouthuysen (FW) transformation and retaining terms up to
$O((|{\bold q}|/M)^{3}$) (M is the nucleon mass
and ${\bold q}$ is the momentum transfer), since the momentum transfer
involved in nuclear scattering of neutrinos produced by accelerators
can be large. The expression for  ${\it H_{eff}}$ can be found in 
\cite{Kubo}.
If the nuclear recoil effects are ignored, we have
\begin{equation}
\sigma={G^{2} \over {2 \pi}}cos^{2}\theta_C\sum_{f}p_lE_l
\int_{-1}^{1}d(cos \, \theta)M_{\beta},
\label{e:2}
\end{equation}
where $G \,cos \, \theta_C$ is the weak coupling constant, $\theta$ is
the angle between the directions of the incident neutrino and the outgoing
lepton and $M_{\beta}$ is given by 
\begin{equation}
M_{\beta} \equiv 
M_{F} \vert  \langle f \vert \tilde 1 \vert i \rangle \vert^2 +
M_{G0} {1 \over 3}
\vert  \langle f \vert \tilde \sigma \vert i \rangle \vert^2 +
M_{G2} \Lambda
\label{e:3}
\end{equation}
where the
squared nuclear matrix elements are
\begin{equation}
\vert  \langle f \vert \tilde 1 \vert i \rangle \vert^2=
{4 \pi \over (2 J_i +1)}\sum_l \vert \langle {\it J_f \Vert 
 \sum_{k} t_{+}(k) j_l(qr_k)Y_l(\hat{{\bold r}}_k) \Vert J_i } \rangle
\vert^2,
\label{e:4}
\end{equation}

\begin{equation}
\vert  \langle f \vert \tilde \sigma \vert i \rangle \vert^2=
{4 \pi \over (2 J_i +1)}\sum_{l,K} \vert \langle {\it J_f \Vert 
 \sum_{k} t_{+}(k) j_l(qr_k)[Y_l(\hat{{\bold r}}_k) 
 \times {\bold \sigma}]^{(K)} \Vert J_i } \rangle
\vert^2,
\label{e:5}
\end{equation}

\begin{eqnarray}
\Lambda & \equiv &  ({5 \over 6})^{1 \over 2}  
\sum_{l,l',K} (-1)^{l/2-l'/2+K} \sqrt{2l+1} \sqrt{2l'+1} {l \, l' \, 2 \choose
0 \, 0 \, 0} 
\left\{1 \, 1 \, 2 \atop l' \, l \, K \right\} {4 \pi \over (2 J_i+1)}
\nonumber \\
 & \times& \langle {\it
J_f \Vert \sum_{k} t_{+}(k) j_l(qr_k)[Y_l(\hat{{\bold r}}_k) 
 \times {\bold \sigma}]^{(K)} \Vert J_i } \rangle 
\nonumber \\
& \times & \langle {\it J_f \Vert \sum_{k'} t_{+}(k') j_{l'}(qr_{k'})
  [Y_{l'}(\hat{{\bold r}}_{k'}) 
 \times {\bold \sigma}]^{(K)} \Vert J_i } \rangle^{\dagger}. 
\label{e:6}
\end{eqnarray}

\noindent
where $k$ labels the space and spin-isospin coordinates of the $k$-th nucleon,
$l,~l'$ are the orbital angular momenta and $K$ is the total angular momentum of
the transition operators.
The coefficients $M_{F}, M_{G0}$ and $M_{G2}$ \cite{Kubo}
appearing in (\ref{e:3})
depend on the momentum transferred to the nucleus
($q=({\bold q},iq_{0})=p_l-p_{\nu}$) and the standard nucleon form factors
${\it f_V(q^2), f_A(q^2), f_W(q^2), f_P(q^2)}$. Second-class current form
factors are ignored.

A correction to (\ref{e:2}) must be introduced to
account for the distortion
of the outgoing lepton wave function due to the
Coulomb field of the daughter nucleus.
For reactions on $^{12}C$
with neutrinos from the
DAR of $\mu^{+}$, the quantity $p_l \, R_A$ ($R_A$ is the radius of the nucleus)
is of the order of 0.5.
In this case, the
cross section (\ref{e:2}) may be multiplied by the Fermi function
$F(Z_{f},E_{l})$ \cite{Macefield},
where $Z_{f}$ is the charge of the daughter nucleus and 
$E_{l}$ the energy of the charged lepton.
When the neutrinos come from  the DIF of $\pi^{+}$, 
the outgoing muons have $p_lR_A > 0.5$. With relativistic leptons, the 
effect due to the Coulomb field
may be included by using the ``Effective Momentum Approximation''
(EMA) \cite{Engel}. In this approximation
the lepton energy and momentum
are modified by 
a constant electrostatic potential within the nucleus, i.e.
$E_{l,eff}=E_l-V(0)$ and $p_{l,eff}=(E_{l,eff}^{2}-m^2)^{1/2}$ with
$V(0)=-3Z_{f}\alpha/2R$.
The cross section (\ref{e:2}) has then to be multiplied by a factor 
$p_{l,eff}E_{l,eff}/(p_lE_l)$.

To obtain the flux-averaged cross sections
$\langle \sigma \rangle_f$ that can be compared with
the experimental data, the energy dependent cross section
(\ref{e:2}) has to be folded with the normalized neutrino
flux $\tilde f(E_{\nu})$ (depending on the neutrino source used)

\begin{equation}\label{e:7}
\langle \sigma \rangle_f = \int dE_{\nu} \sigma(E_{\nu}) \tilde f(E_{\nu}),
\end{equation}

where 
\begin{equation}\label{e:8}
\tilde f(E_{\nu})= { f(E_{\nu}) \over {\int_{E_0}^{\infty} 
  dE^{'}_{\nu} f(E^{'}_{\nu})}},
\end{equation}

\noindent
$f(E_{\nu})$ being the initial flux and
$E_{0}$ 
the threshold energy. 

Closely related to this neutrino
capture reaction is the capture of a negative muon
bound in an atomic orbit, $\mu^- + (A,Z) \rightarrow (A,Z-1)^* + \nu_{\mu}$.
In the 1S-capture the inclusive rate 
$\Lambda_c$
 is given by
\begin{equation}\label{e:1mucap}
\Lambda_c=
{m_{\mu}^2 \over 2 \pi}
\vert \phi_{1S} \vert^2
[G_{V}^2M_V^2+G_{A}^2M_A^2+(G^2_P-2G_PG_A)M_P^2].
\end{equation}
if we neglect
the recoil term
which represents a correction of a few percent \cite{AK84}.
The function $\phi_{1S} $ is the muon 
1S-bound state wave function evaluated at the origin,
i.e., $\vert \phi_{1S} \vert^2= R(Z \alpha m')^3/\pi$,
$R$ being a reduction factor accounting for the finite size of
the nuclear charge distribution ($R=0.86$ for $^{12}C$) \cite{Uber72}
and $m'$ the muon reduced mass.
The constants $G_V,G_A,G_P$ \cite{Luy} are the
``effective coupling constants'' which
depend only slightly on the neutrino momentum $p_{\nu}$.
This can be simply obtained
from the energy and momentum conservation,
$p_{\nu}=m' - (m_n - m_p)- \vert E^{B}_{\mu} \vert - E_{fi} $, where
$m_n,m_p$ are the neutron and proton masses,
$\vert E^{B}_{\mu} \vert $ is the binding energy of the muon in
the $1S$ orbit and $E_{fi}$ is the nuclear excitation energy
measured with respect to the parent nucleus ground state.
The capture rate can be factorized as in (\ref{e:1mucap}) if we 
neglect the dependence of the coupling constants on $p_{\nu}$.
The square of the vector, axial-vector and pseudoscalar matrix elements
are 
\begin{equation}\label{e:2mucap}
M_V^2=4 \pi \sum_l (2l+1) \sum_f ({p_{\nu} \over m_{\mu}})^2
\vert   \langle {\it J_f \vert \sum_k t_+(k) 
j_l(p_{\nu} r_k)Y_{l0}(\hat{\bf r}_k)\vert J_i } \rangle \vert^2,
\end{equation}
\begin{equation}\label{e:3mucap}
M_A^2=4 \pi \sum_{l,K} (2K+1) \sum_f ({p_{\nu} \over m_{\mu}})^2
\vert   \langle {\it J_f \vert 
  \sum_k t_+(k) j_l(p_{\nu} r_k)[Y_{l}({\bf \hat{r}}_k) \times \sigma]^{K0} 
\vert J_i } \rangle \vert^2,
\end{equation}

\begin{equation}\label{e:4mucap}
M_P^2=4 \pi \sum_{l,l',K} (2K+1) {\it \sum_f ({p_{\nu} \over m_{\mu}})^2
\vert   \langle  J_f \vert \sum_k t_+(k) 
\theta_P(K)
\vert J_i }  \rangle \vert^2, 
\end{equation}

where

\begin{equation}
\theta_P(K)=
\left\{
\sqrt{K \over (2K+1)}j_l(p_{\nu} r_k)[Y_{l}({\bf \hat{r}}_k) 
\times \sigma]^{K0}  +  
\sqrt{(K+1) \over (2K+1)} 2 j_l'(p_{\nu} r_k)[Y_{l'}({\bf \hat{r}}_k) 
\times \sigma]^{K0} 
\right\}. \nonumber
\end{equation}
with $l=K-1$ and $l'=K+1$.
The $\beta$-decay corresponds to the limit of zero momentum transfer
of the
transition probabilities in (\ref{e:3}), that is 
\begin{equation}\label{e:1beta}
M^0_{\beta} \equiv 
f_{V}^2(0) \vert  \langle f \vert 1 \vert i \rangle \vert^2 +
f_{A}^2(0) {1 \over 3}
\vert  \langle f \vert \sigma \vert i \rangle \vert^2 
\end{equation}
where 
\begin{equation}\label{e:2beta}
\vert  \langle f \vert 1 \vert i \rangle \vert^2=
{1 \over{2J_i+1}}\vert \langle 
{\it J_f \Vert \sum_k t_+(k)\Vert J_i }\rangle \vert^2
\end{equation}

\begin{equation}\label{e:3beta}
\vert  \langle f \vert  \sigma \vert i \rangle \vert^2=
{1 \over{2J_i+1}} \vert \langle {\it J_f \Vert \sum_k t_{+}(k)
{\bold \sigma}(k)  \Vert J_i}\rangle \vert
\end{equation}
and the transition operators are of the usual
Fermi or Gamow-Teller type.
The {\it ft} value is given by
\begin{equation}\label{e:4beta}
{\it ft}= {2 \pi^3 ln2 \over (G^2 cos^2\theta_C m_e^5)} {1 \over  M_{\beta}^0}.
\end{equation}

The neutrino reaction cross section (\ref{e:2}),
the muon capture rate (\ref{e:1mucap}) and the {\it ft} value for the
beta decay (\ref{e:1beta}) depend on the
wavefunctions of the initial and final
nuclear states involved in these processes. 
The microscopic models used in this work 
to evaluate these wavefunctions and the corresponding
transition probabilities are described in the next section.

\section{Microscopic models}
\subsection{RPA and QRPA}
Transition matrix elements of the type entering in eqs.
(\ref{e:3},\ref{e:1mucap},\ref{e:1beta}) 
can be calculated within the framework of RPA or QRPA. In the
present work, the starting point is a Hartree-Fock (HF) calculation of the
ground-state of $^{12}$C, performed in coordinate space by using the 
Skyrme-type effective interactions SIII~\cite{Bei75} and SGII~\cite{Gia81}. 
The SGII force was built with the purpose of obtaining a proper 
description of spin-isospin nuclear properties. The HF solution determines 
the mean-field and single-particle (s.p.) occupied levels of $^{12}$C which 
has a closed-subshell structure in this description. The unoccupied levels of $^{12}$C 
are obtained by diagonalizing the HF mean-field using a harmonic oscillator basis.
Therefore, the continuum part of the s.p. spectrum is discretized and discrete 
particle-hole (ph) configurations coupled to $J^\pi$ are used as a basis in 
order to cast the RPA equations in the matrix form. The details of this 
procedure can be found in~\cite{Col94}. This RPA calculation is 
self-consistent since the residual interaction among ph states is derived 
from the same Skyrme force used to produce the mean field.

To go beyond the closed-subshell approximation for the $^{12}$C ground state,
pairing correlations are taken into account in the 
HF+BCS approximation. 
Constant pairing gaps $\Delta_p$ and $\Delta_n$ for protons
and neutrons are introduced and are set 
 at 4.5 MeV. 
 A large pairing gap is unrealistic for
states far from the Fermi surface, and an energy cut-off  is required, such that
the states above this cut-off have $\Delta$ = 0. The cut-off is set at
the $2s_{1/2}$ state.
On top of the HF+BCS calculation, the QRPA matrix equations can be written 
with a procedure which parallels what was described
above, with the two-quasiparticle 
(2qp) configurations replacing the ph ones. We do not present here the details 
of the QRPA formalism, which is found in the literature (see, e.g., 
Ref.~\cite{Gro90}). We simply note that the particle-particle matrix elements 
are here renormalized by means of a parameter $g_{\rm pp}$ that has been 
chosen to be smaller 
than 1 (typically 0.7) to avoid the well-known ground-state 
instabilities. 

For the multipolarities studied, it is found that
the space used for the RPA (QRPA) calculations
satisfies well (up to a few percent)   the 
energy-weighted sum rule (EWSR) associated with the 
operators of the type
\begin{equation}\label{e:SMO}
\sum_{k} t_{+}(k) r_k^l Y_l(\hat{{\bold r}}_k), \ \ \ 
\sum_{k} t_{+}(k) r_k^l [Y_l(\hat{{\bold r}}_k) \times {\bold \sigma}]^{(J)}
\end{equation}
which are the small-$q$ limit of those defined in the preceding section.

For each multipolarity,
every eigenstate of the
RPA or QRPA equations is characterized by its $X^f$ and $Y^f$ amplitudes
and the transition matrix element for
a generic operator $\hat{O}(k)$ is written as
\begin{equation}
\langle {J_f \vert \sum_k \hat O(k) \vert \it J_i} \rangle =
\sum_{\alpha,\beta} \langle \alpha \vert \sum_k \hat O(k) \vert \beta \rangle
(X^f_{\alpha\beta}u_\alpha v_\beta + Y^f_{\alpha\beta}v_\alpha u_\beta), 
\end{equation}
where $\alpha$ and $\beta$
label a given ph or 2qp states, $u$ and $v$ are the
BCS occupation amplitudes
(which reduce to 1 and 0 in the HF-RPA case) and
$\langle \alpha \vert \sum_k \hat O(k) \vert \beta \rangle$ are single-particle
matrix elements.
With $\Delta=4.5~MeV$, the occupation probabilities for the $1p_{3/2}$ and
$1p_{1/2}$ states are $v^2_{p_{3/2}}=0.84$ and $v^2_{p_{1/2}}=0.19$ 
respectively. The latter one 
is smaller than the one used in \cite{Vorov,KL99}.


\subsection{Shell Model}
We also evaluate transition matrix elements for neutrino and muon
capture reactions using a large shell model space.
It is desirable to use extended spaces as much as possible in order to treat
both inclusive and exclusive reactions.
Here, we take the {\it 0s-0p-1s0d-1p0f} shell model space and include
configurations 
up to $3 \hbar\omega$ excitations 
for negative parity states and up to 2$\hbar\omega$ excitations
for positive parity states.  No $^{4}$He core is
assumed in the present calculations.  The configurations taken for
positive parity states in A=12 nuclei are
\begin{eqnarray}
(0s)^4(0p)^8 &+& (0s)^4(0p)^6(1s0d)^2 + (0s)^4(0p)^7(1p0f)^1\nonumber\\
 &+& (0s)^3(0p)^8(1s0d)^1 + (0s)^2(0p)^{10}
\end{eqnarray}
and those for negative parity states are
\begin{eqnarray}
(0s)^4(0p)^7(1s0d)^1 &+& (0s)^3(0p)^9 + (0s)^4(0p)^6(1s0d)^1(1p0f)^1  
\nonumber\\
&+& (0s)^4(0p)^5(1s0d)^3 + (0s)^3(0p)^8(1p0f)^1\nonumber\\
&+& (0s)^3(0p)^7(1s0d)^2 + (0s)^2(0p)^9(1s0d)^1
\end{eqnarray}
In the shell model calculations of ref.\cite{Hayes} of the present problem, 
computations
were carried out with configurations up to 1$\hbar\omega$ excitation
for negative parity states and up to 2$\hbar\omega$ excitation
for positive parity states.

The spurious center-of-mass states are eliminated here by using the method
of Lawson\cite{Lawson}. The spurious states are pushed up to higher
energies with the addition of a fictitious term in the
hamiltonian which acts only on
the center-of-mass excitations.
The number of spurious states removed is $\sim 1000-4000$ for each negative
parity multipole ($0^{-}$, \dots, $4^{-}$) and $\sim 100-400$ for each
positive parity multipole ($0^{+}$, \dots, $6^{+}$).  The number of
states (without the spurious center-of-mass components) amounts to 
$\sim 2800-10600$ for the negative parity multipoles and $\sim 250-1500$
for the positive
parity multipoles. The number of states
is very large for the
negative parity multipoles compared to the work in ref.\cite{Hayes},
in which the number was $\sim 50-150$ and included 
up to 1$\hbar\omega$ excitation only.

We adopt here the effective interaction of Warburton and Brown\cite{WB}
for use in the present {\it 0s-0p-1s0d-1p0f} model space, and 
we use the
set WB10\cite{OXB}, which is based on
the WBT interaction\cite{WB}.
This interaction was obtained by fitting binding energies and energy
levels, including cross-shell data.  The interaction describes well 
the low excitation energy spectra for A= 10-22 nuclei.
It has also been used to investigate Gamow-Teller $\beta$ decay rates
for A $\leq$18 nuclei\cite{CWB}.  

In the shell model, the reduced matrix elements of transition operators 
are expressed as linear combinations 
of the reduced matrix elements of single-particle
states with coefficients given by one-body density matrix elements
\begin{equation}
\langle J_{f} T_{f} || \hat{O}^{\lambda t} || J_{i} T_{i}\rangle
= \sum_{j_{i}j_{f}} C_{j_{i}j_{f}}^{\lambda t} \langle j_{f} || 
\hat{O}^{\lambda t} || j_{i}\rangle.   
\end{equation}
The form factors in (\ref{e:3}) have to be corrected for
the center-of-mass motion. This is done by multiplying
the matrix elements by the Tassie-Barker function, 
$exp(b^2q^2/2A)$, with $b$ being the oscillator
length parameter \cite{Tassie}.  

\section{Discussion of the results}
\subsection{Theory versus experiment}
Using the above formalism we calculate results for the flux-averaged
cross sections for neutrinos coming from the DIF of $\pi^{+}$,
$(\nu_{\mu},\mu^-)DIF$, and for neutrinos coming from the DAR of $\mu^{+}$,
$(\nu_{e},e^-)DAR$.
The 
flux-averaged
cross sections are obtained by taking neutrino fluxes from \cite{Imlayfl}.
As far as RPA and QRPA are concerned, we discuss in the following
results obtained by using the force SIII. We have checked that
the interaction SGII gives very similar results.

Let us first discuss our results for the inclusive cross sections
(table \ref{tab:1}).
In the Shell Model (SM) case, cross sections obtained both in the
{\it 0s-0p-1s0d}
((0+1+2)$\hbar \omega$) model space and in the
{\it 0s-0p-1s0d-1p0f} ((0+1+2+3)$\hbar \omega$) model space
are shown. Results in the smaller space are only given for comparison
with \cite{Hayes}. Two different types of radial wavefunctions have been used,
the Harmonic Oscillator  with $b=1.64~fm$ (HO wf) and Hartree-Fock
wavefunctions (HF wf). HO wavefunctions are used even though their radial
behaviour is known not to give good results of $^{12}C(p,n)$ reactions
cross sections.
The reason for using them is twofold :
i) the spurious center-of-mass motion is exactly substracted
only in this case; ii) they are employed
in order to show the sensitivity of the calculated
cross sections to the
choice of the radial wavefunctions.
Cross sections can be indeed quite sensitive to the radial
wavefunctions since going from  HO to HF wavefunctions in the
$(\nu_{\mu},\mu^-)DIF$ case produces changes of about $30 \%$, whereas in the
$(\nu_{e},e^-)DAR$ case we have variations of a few percent only
(see table \ref{tab:1}).
For comparison we also show the results of ref.\cite{Hayes} 
obtained by extrapolation in the {\it 0s-0p-1s0d-1p0f} space
with Woods-Saxon wave functions.
The results obtained within the
Skyrme RPA and QRPA are larger than in SM. 
The QRPA values are sligthly larger
than the RPA ones
for the inclusive cross section.
This is due to the fact that in QRPA the
particle-particle residual
interaction (on the average attractive)
competes with the (on the average repulsive)
particle-hole residual interaction. 
As a consequence, the QRPA
strength distribution is slightly shifted towards lower energies.
This fact produces a few percent increase of the flux-averaged cross sections
due to the dependence of the differential cross sections (\ref{e:2}) on
the energy and momentum of the outgoing lepton and also due
to the flux averaging
(\ref{e:7},\ref{e:8}). In ref.\cite{Vorov,KL99}, on the
contrary the introduction of fractional occupancies shifts the strength
distribution to higher energies and this gives a few percent decrease
of the flux averaged cross section with respect to their RPA results.
From table \ref{tab:1}, we see that the
$(\nu_{\mu},\mu^-)DIF$ cross sections are compatible with those of
ref.\cite{KL99}, but there is a ratio of about a
factor 3 for the $(\nu_{e},e^-)DAR$ case. 
Most of the disagreement 
is due to the bad description of the ground state to ground state
transition (its contribution represents 2/3 of the total cross section),
some of it arises because of the ground state energy
of $^{12}N$ which is not close to the experimental value. 
In \cite{KL99} the single particle energies were fitted
to the experimental values whereas our single particle energies are
kept as they come out from our Skyrme HF or HF-BCS calculation.
The single particle energies are particularly important to get the energy
of the $^{12}N$ ground state with respect to $^{12}C$ ground state close to
its experimental value of $17.338~MeV$. In fact, within
RPA this transition
corresponds to the $1p_{3/2} \rightarrow 1p_{1/2}$ transition,
and with the Skyrme forces employed in this work the energy of
this transition is $13.4~MeV$. 

In table \ref{tab:2}, we see the importance of including configuration mixing 
as well as 
of having the energy of this transition close to its experimental value.
This can also be
seen from table \ref{tab:3}, where we give 
the inclusive $(\nu_{e},e^-)DAR$ cross sections
calculated without the contribution of the ground state to ground state 
transition. In this case, our 
RPA cross section is very close to the experimental value as found in
\cite{Vorov,KL99}. 

We discuss now the
theoretical predictions versus the
experimental findings. For the $(\nu_{\mu},\mu^-)DIF$ case, we see 
from table \ref{tab:1} that
our results within SM (with HF wf) are close to the experimental
value when the error bars are taken into account, sligthly overestimating it.  
The WS results in \cite{Hayes} agree even better. 
On the other hand, the RPA, the QRPA and
Continuum RPA 
(CRPA) with fractional occupancies  \cite{KL99} overestimate the experimental value by about $50 \%$.
Concerning the $(\nu_{e},e^-)DAR$ cross sections, 
we see that the SM results
({\it 0s-0p-1s0d-1p0f} with HF wf) are compatible with the experimental values
when error bars are taken into account, as those of ref.\cite{Hayes}. 
Again the CRPA results of ref.\cite{KL99} overestimate the experimental values
by about $30 \%$ (while our RPA predictions are far off for the reasons
explained above).

Let us discuss the exclusive cross sections together with various
related processes like $\beta$ decay and muon capture.
First, we can see from table \ref{tab:2} that 
our shell model results nicely agree with the
experimental values for both the DIF and the DAR cases. The calculated values 
are not very sensitive to the choice of the radial wave functions and confirm 
those calculated in \cite{Hayes}. 
Our RPA prediction for the $(\nu_{\mu},\mu^-)DIF$
is about a factor 3 larger than the experimental value as found also in
\cite{KL95},
whereas the introduction of fractional occupancies brings
the RPA predictions close  to the experimental value \cite{Vorov,KL99}.
As far as the QRPA is concerned, the value obtained is slightly
lower than in RPA, but within QRPA there are difficulties in choosing the
ground state of $^{12}N$ because the lowest state
is not the most collective one. Therefore all the results concerning the
exclusive transition have been obtained by summing the strength of the
energy levels within the first $3~MeV$ above the lowest state
and attributing it to a single state at an average energy
which is not much different from the RPA one.
Our RPA $(\nu_{e},e^-)DAR$ exclusive cross section results 
are a factor 4-5 larger than
the experimental values as found in \cite{KL95}.
A related process is the $^{12}N \rightarrow ^{12}C$
$\beta^+$ decay.
In this case the transition operator does not have a radial dependence.
From table \ref{tab:4}, we see that the SM 
{\it ft} value is in reasonable agreement with 
the experimental one whereas RPA gives values
four times smaller as it is well known \cite{KL95,KL99}.
The {\it ft} value for the $\beta^-$ decay from $^{12}B$ to $^{12}C$ is not
given, but similar arguments hold since $^{12}B$ and $^{12}N$ are mirror
nuclei. 
Finally, let us look at the exclusive muon capture rates (table \ref{tab:5}).
The SM results are very close to the experimental value in agreement with
what was found in \cite{Hayes} while as in \cite{KL99} RPA overestimates it by
about a factor 4.
The nice agreement between experiment and theory obtained within SM 
(at variance with standard RPA) shows
clearly that the inclusion of configuration mixing is necessary for a
good description of the mentioned processes.

We now turn to the discussion of the inclusive $(\nu_{e},e^-)DAR$
cross section without the contribution of the ground state to ground state
transition (see table \ref{tab:3}).
There are two main differences with the results in \cite{Hayes} :
i) our $(\nu_{e},e^-)DAR$ cross sections are not very sensitive
to the choice of the wavefunctions;
ii) the calculated value is twice the one obtained in \cite{Hayes}. 
In order to understand point i), we show
the strength distribution obtained for
the multipolarity $J^{\pi}=1^+$, calculated either with
HO or HF wavefunctions, 
with  $q=0.2~fm^{-1}$ in
(\ref{e:5}) as a ``typical'' value of the momentum transferred
in the $(\nu_{e},e^-)DAR$ reaction (fig.1)
and $q=1.0~fm^{-1}$ as a ``typical'' value
for the $(\nu_{\mu},{\mu}^-)DIF$ case (fig.2). 
We see that the strength distributions 
obtained for $q=0.2~fm^{-1}$ are the same for the two s.p. wavefunctions,
while for $q=1.0~fm^{-1}$
there are significant differences.
This explains the sensitivity to the choice of the wavefunctions
of the results of table \ref{tab:1} and \ref{tab:3}. 
Concerning point ii), from (\ref{e:2}) we see that
the differential cross sections scale as the square of the lepton energy
in the case of electrons (because of its negligeable mass).
As a consequence, small shifts in the strength distributions can have
large effects on the cross sections as we have already seen in table
\ref{tab:2}.
Because neutrino-nucleus reactions with electronic neutrinos are
so sensitive to the details of the strength distribution,
they could eventually be used as a probe 
of nuclear structure.

Inclusive muon capture rates, calculated according to
(\ref{e:1mucap})-(\ref{e:4mucap}) are given in table \ref{tab:6}.
The SM results in the large space and with HF wavefunctions are
about $20 \%$ lower than in \cite{Hayes} and $10 \%$ lower than the
experimental values. Concerning the results obtained within RPA,
the disagreement between calculations 
(present results and ref.\cite{KL99})
and the experimental value is again due to the bad description of the ground
states, as it can be seen from table \ref{tab:5}.

The contributions of the most important multipolarities with
$J \le 6$ included in the inclusive $(\nu_{\mu},{\mu}^-)DIF$ 
and $(\nu_{e},e^-)DAR$ cross sections are shown
in tables \ref{tab:7} and \ref{tab:8} respectively.
We see that the contributions in RPA and QRPA are essentially the same,
showing that the configuration mixing induced by the pairing correlations
is not affecting very much the
inclusive cross sections. 
In table \ref{tab:7} the results of the SM calculations again depend significantly
on the choice of the s.p. wavefunctions. Significant differences
are also evident going from RPA to SM (HF wavefunctions), for example
for the $1^+,~1^-,~2^+$ and $3^-$. On the contrary, the contributions of the
different multipolarities to the $(\nu_{e},e^-)DAR$
cross sections are almost equal in the three approaches, 
except for the remarkable difference associated with
the $1^+$ states (and in particular with the ground state to
ground state contribution).

In view of future experiments using $^{12}C$ as detector for neutrinos,
with impinging neutrino fluxes different from the ones used up to now,
in figs.3 and 4 we give the differential cross sections obtained
both in RPA and in SM for the reactions
$^{12}C(\nu_e,e^-)^{12}N$ and $^{12}C(\nu_{\mu},{\mu}^-)^{12}N$
as a function of neutrino energies up to $300~MeV$.

A question arises concerning the reason for the difference between
the RPA and SM results for the inclusive cross
section in the $(\nu_{\mu},{\mu}^-)DIF$ case
(tables \ref{tab:1},\ref{tab:7}), 
the value of which has important implications
on the recent measurements of neutrino
oscillations.
The two approaches have two major differences :
i) the type of correlations included;
ii) the model space used.
Concerning point i), we have already seen in the previous section that
correlations
are actually responsible for the quenching
in the $1^+$ (ground state to ground state) contribution (table
\ref{tab:7}) which is taken into account more correctly in the SM.
Due to the tensor interaction
the effects of quenching and fragmentation of the strength are also
important for the spin-dipole mode, especially for the $1^{-}$
component 
(see the discussion in ref.\cite{SS}).
The difference of cross sections between SM and RPA in the case
of the $0^-$ and $2^-$ is rather small, while it is about
$0.5 \times 10^{-40}~cm^{2}$ for the $1^-$. 
The difference in the type of the correlations included
explains half of the discrepancy between the SM and RPA results.
Concerning ii),
one important constraint on nuclear models
is due to sum rules \cite{Lip}. 
The existence of sum rules
imposes constraints on the model spaces to be used in a given approach
because one should ensure that the model space used is large
enough to satisfy the corresponding sum rule for a given operator.
This has not been considered in previous studies \cite{KL95,Hayes,KL99}
and can explain some of the difference between the RPA and SM cross sections.
In the next subsection we discuss the role of sum rules. 

\subsection{Sum rules}
The processes studied
involve operators of two kinds, either of Fermi type (\ref{e:4})
or of Gamow-Teller type (\ref{e:5}). In the low-$q$ limit
these operators become the standard multipole operators 
(\ref{e:SMO}). 
As an example we have calculated sum rules for the 
non spin-flip states
$J^{\pi}=1^-$ and $J^{\pi}=2^+$  both in RPA and in SM.
We have found that if for the $1^-$ states, the corresponding sum rule
is satisfied in the two approaches, the sum rule for the $2^+$ states is
satisfied in the RPA whereas $20 \%$  is missing in SM when the largest space
({\it 0s-0p-1s0d-1p0f}) is used. Figure 5 shows a comparison between
the RPA and SM strength distributions obtained for the $2^{+}$ states
with the Fermi type operator, $\hat{O}=\sum_k r_k^2Y_2(r_k)t_{+}(k)$. 
We can see that in the SM calculation
some strength is missing at about $40-60~MeV$ because of the truncation
of the space.
This missing strength at high energy gives a significant contribution to
the flux averaged $(\nu_{\mu},{\mu}^-)DIF$ cross section. 

In order to show this we have performed an RPA calculation of this
cross section in the same space as used in the SM calculations
({\it 0s-0p-1s0d-1p0f}).
The results obtained in RPA,
RPA in the restricted ($3 \hbar \omega$) model space and SM
are shown in table \ref{tab:9}. We see that
 for the $1^-$ states, for which the basis
is large enough for
the sum rule to be satisfied, the corresponding 
cross sections are practically the same
in the three cases.
On the contrary, the $2^{+}$ contributions to the total cross section
look quite different, showing a reduction of about $10 \%$ going from RPA
to RPA in the restricted space and to almost $30 \%$ in the SM.
This reduction is due to the missing strength 
as illustrated in figure 5 and is
accompanied by sum rules that are not exhausted.

Finally, from table \ref{tab:9} we also see that restricting
the RPA space brings the total flux averaged cross section very close
to the SM one. This important result seems to indicate strongly that getting
theoretical predictions in the shell-model framework which
come close to the experimental values in the inclusive
$(\nu_{\mu},{\mu}^-)DIF$ can be 
an artifact because the model spaces used for the calculations are
not large enough to satisfy sum rules.
When a more extended space is used, it is quite possible that the shell
model cross section will increase approaching
the RPA value and exceeding the experimental
LSND result.
However, we should also note that the increase of cross sections with 
increasing configuration space is not the same for SM and
RPA.  The rate of increase of cross sections is smaller for
the SM case.  This can be already seen from the difference of the rate 
of increase when going from $1\hbar\omega$ to 
$3\hbar\omega$ spaces when we compare our SM results with
ref.\cite{Hayes}.
The rate of increase is $14\%$ in our case while it is $19\%$ 
in ref.\cite{Hayes}, where the cross section for $3\hbar\omega$ space
was obtained from an extrapolation based on an RPA-like model.

\subsection{The problem of quenching of spin strength}
Finally, we would like to discuss more the problem of quenching of spin
  strength in nuclei.
It is known for sometime that there is not enough strength in the main
GT peak.
Two explanations on the origin of this quenching have been suggested :
i) the missing strength is shifted to very high energy
due to the coupling of the nucleon internal structure to the Delta
resonance; ii) there is a shift of the strength to energies
up to about, or more than $50~MeV$
due to the coupling to multi-particle multi-hole configurations.
The quenching and fragmentation of the spin strength due to these
effects would lead to quenching of the cross sections of the GT mode
and probably for the spin-dipole modes as well.
We have made a test calculation in which we have
introduced
an effective axial vector
coupling constant, $g_{A}^{eff}$.
Considering high energies of neutrinos in the DIF
experiments most of the missing strength
due to the above
multiparticle effects is picked up
and also some portion of the strength
due to the coupling to the Delta-Resonance effect is recovered 
\cite{Arima,Towner}.
Therefore the $g_{A}^{eff}$ one should use here should be
closer to $g_{A}$ as compared to the value found for the main GT
peak ($g_{A}^{eff}= 0.8g_{A}$).


In table \ref{tab:10}, 
we show all the results concerning the different processes
using a quenching factor.
For the sake of demonstration we use $g_{A}^{eff}=0.9g_{A}$.
Using this value, the {\it ft} value is increased as expected by $20 \%$ and
the exclusive and inclusive muon capture is reduced by about $15-20\%$.
Concerning 
the exclusive cross sections are reduced
by about $15 \%$ and become smaller than the observed ones, 
whereas the inclusive ones are reduced by about $10-15 \%$ 
and get closer to the experimental values.  
The exclusive neutrino cross sections we have seen are very sensitive
to the calculated wavefunctions and therefore cannot be a good test
of the quenching factor. The inclusive cross sections are less sensitive to
nuclear structure and they seem to favor some reduction in $g_A$. There
are many uncertainties in these arguments. We do not yet know for sure
what is the mechanism of quenching of $g_A$ 
(although the experiment in \cite{Sakai} 
suggests that the main role is played by the coupling with many 
particle-many hole states)
and whether
the same quenching factor should be applied to all multipolarities.
It is quite conceivable that future studies of the neutrino-nucleus
interaction will actually help to clarify this point about quenching.

\section{Conclusions}
Microscopic approaches, namely charge-exhange RPA, charge-exhange QRPA and Shell Model,
are used to evaluate $^{12}C(\nu_e,e^-)^{12}N$ and
$^{12}C(\nu_{\mu},{\mu}^-)^{12}N$ cross sections
both for $\nu_e$ coming from the decay-at-rest of $\mu^+$ and
for $\nu_{\mu}$ coming from the decay-in-flight of $\pi^+$.
Accurate knowledge of these cross sections is important for
the interpretation of recent measurements of neutrino oscillations
performed both by LSND and KARMEN collaborations and also for future
experiments.
The results show that
the calculated exclusive cross sections, where $^{12}N$ is left in the
ground state, are in good agreement with the measured values when
large-scale shell model calculations are performed.
In fact, in this framework it is easy
to  include properly the configuration mixing present in the ground
state of $^{12}C$ and therefore to have a good description of the
ground state wavefunctions.
Concerning the inclusive $^{12}C(\nu_{\mu},{\mu}^-)^{12}N$
cross sections with $\nu_{\mu}$ from DIF of $\pi^+$, in which $^{12}N$ is left
either in the ground state or in an excited state, we get
in the shell model 
$15.2 \times 10^{-40}~cm^2$
(when Hartree-Fock
wavefunctions are used),
about $20 \%$ larger than the experimental value, which is
$12.4 \pm 0.3 \pm 1.8 \times 10^{-40}~cm^2$.
The calculated 
cross section in charge-exchange RPA is $50 \%$ larger than this value,
namely $19.2 \times 10^{-40}~cm^2$.
The most important difference between the SM and RPA is the quenching
in the $1^{+}$ due to the different correlations included in the
two approaches. 
This explains
half of the difference between the two results.   
The evaluation of sum rules for natural parity states
has also shown that
the basis used in the shell model calculations is not
large enough to exhaust these sum rules 
whereas in the case
of the RPA the sum rules are satisfied. 
This fact suggests that 
the reduced cross section obtained in the shell model
framework is partially due to the use of a basis which is not large enough.
Enlarging the model space would then add some strength at high energy
 and therefore
increase the inclusive cross sections.
From these arguments, we conclude that in both microscopic
approaches the theoretical prediction is $20-30 \%$ larger than
the measured value. But even if we could hope to extend further the
shell model calculation and increase the basis, the calculated value
will keep having $10-20 \%$ uncertainty due to a certain degree
of arbitrariness in the choice of the wavefunctions for the unbound states
and also of the interactions. 
It would be interesting
to extend the shell model calculations to larger 
configuration space in spite of the uncertainties above,
to have more information on the correlations in spin-dipole 
modes and clarify the convergence of the cross sections as the space
is extended. 
Concerning the inclusive $^{12}C(\nu_e,e^-)^{12}N$ cross section we
get $16.4 \times 10^{-42}~cm^2$
which agrees within the experimental error bars with the 
measured value.
The $ft$ value for the beta decay from $^{12}N$ to $^{12}C$,
exclusive and inclusive muon capture rates $^{12}B(\mu^{-},\nu_{\mu})^{12}C$
are also evaluated and are in quite good agreement with the measured values.

This work was supported by the US-Israel Binational Science Foundation.
The shell model calculation was done by using the TKYNT 
at Department of Physics, University of Tokyo.

\newpage

\begin {table}
\caption{Flux averaged inclusive cross sections $<\sigma>_{f}$ within
the different approaches used, namely the Shell Model (SM), the Random Phase
Approximation (RPA) and the Quasi-particle RPA (QRPA).
Different SM results are given according to
various choices of radial wavefunctions, i.e.,
oscillator functions with length parameter $b=1.64~fm$ (HO wf)
 and Hartree-Fock
wave functions (HF wf) and for
either $(0+1+2)\hbar \omega$ or $(0+1+2+3)\hbar \omega$ model space. 
Comparison with CRPA with fractional
occupancies \protect\cite{KL99} 
and a recent shell model calculation \protect\cite{Hayes} is
made. In the former case the results are obtained with the
finite-range G-matrix derived from the Bonn NN potential (BP)
and with the Landau-Migdal (LM) force (in brackets).
In the latter case, Woods-Saxon wavefunctions (WS wf) have been used
and the results within the $(0+1+2+3)\hbar \omega$ model space (in brackets) 
are obtained by extrapolation.}
\begin{center}

\begin{tabular}{lcc} 
  & $(\nu_{\mu},\mu^-)DIF$  & $(\nu_{e},e^-)DAR$ \\
  & $<\sigma>_{f} (10^{-40}~cm^2)$ & $<\sigma>_{f} (10^{-42}~cm^2)$  
  \\ \hline 

SM(HO wf) $(0+1+2)\hbar \omega$  & 18.71 & 14.21 \\
SM(HF wf) $(0+1+2)\hbar \omega$  & 13.33 & 13.94  \\
SM(WS wf) $(0+1+2)\hbar \omega$  \cite{Hayes}  & 11.1 & 12.1 \\
SM(HO wf) $(0+1+2+3)\hbar \omega$  & 21.08 &  16.70\\
SM(HF wf) $(0+1+2+3)\hbar \omega$   & 15.18  & 16.42  \\
SM(WS wf) $(0+1+2+3)\hbar \omega$  \cite{Hayes}  & $(13.2)$  & $(12.3)$ \\
RPA & 19.23 & 55.10 \\
QRPA & 20.29  & 52.0   \\
CRPA \cite{KL99}  & 18.18(17.80) & 19.28(18.15) \\
EXP  & $12.4 \pm 0.3 \pm 1.8$ \cite{LSNDvmu},\cite{KL99} &
$14.1 \pm 1.6 \pm 1.9 $ \cite{Allen} \\
     &      & $14.8  \pm 0.7 \pm 1.4 $ \cite{LSNDve}          \\
     &      & $14.0 \pm 1.2$ \cite{KAR_ve_Ngs} \\ 
\end{tabular}  
\label{tab:1}
\end{center}
\end {table}

\begin {table}
\caption {Same as table 1 for flux averaged exclusive cross sections.}
\begin{center}

\begin{tabular}{lcc} 
  & $(\nu_{\mu},\mu^-)DIF$  & $(\nu_{e},e^-)DAR$ \\
  & $<\sigma>_{f} (10^{-40}~cm^2)$ & $<\sigma>_{f} (10^{-42}~cm^2)$  
  \\ \hline 

SM(HO wf) $(0+1+2)\hbar \omega$ & 0.70  &  8.42  \\
SM(HF wf) $(0+1+2)\hbar \omega$ & 0.65  &  8.11  \\
SM(WS wf) $(0+1+2)\hbar \omega$ \cite{Hayes}  & 0.58 & 8.4 \\
RPA & 2.09 & 49.47  \\
QRPA & 1.97  & 42.92   \\
CRPA \cite{KL99}  & 1.06(1.03) & 13.88(12.55) \\
EXP  & $0.66 \pm 1.0 \pm 1.0$ \cite{LSNDvmu} & 
$10.5 \pm 1.0 \pm 1.0$ \cite{Allen} \\
     &      & $9.1 \pm 0.4 \pm 0.9$  \cite{LSNDve} \\
     &      & $9.1 \pm 0.5 \pm 0.8$  \cite{KAR_ve_Ngs}  \\ 
\end{tabular}  
\label{tab:2}
\end{center}
\end {table} 

\newpage

\begin {table}
\caption {Same as table 1 for flux averaged 
  inclusive cross sections but excluding 
the ground state.}
\begin{center}

\begin{tabular}{lc} 
  & $(\nu_{e},e^-)DAR$ \\
  & $<\sigma>_{f} (10^{-42}~cm^2)$  
  \\ \hline 

SM(HO wf) $(0+1+2)\hbar \omega$  &  5.79     \\
SM(HF wf) $(0+1+2)\hbar \omega$  &  5.83    \\
SM(WS wf) $(0+1+2)\hbar \omega$ \cite{Hayes}     & 3.7 \\
SM(HF wf) $(0+1+2+3)\hbar \omega$   & 8.28 \\
SM(HO wf) $(0+1+2+3)\hbar \omega$  &  8.31\\
SM(WS wf) $(0+1+2+3)\hbar \omega$ \cite{Hayes}  & $(3.8)$ \\
RPA & 5.63 \\
QRPA & 9.08 \\
CRPA \cite{KL99}  & 5.4(5.6) \\
EXP  & $5.4 \pm 1.9$ \cite{Allen} \\
     & $5.7 \pm 0.6 \pm 0.6 $ \cite{LSNDve} \\
     & $5.1 \pm 0.8$ \cite{KAR_ve_Ngs}         \\ 
\end{tabular}  
\label{tab:3}
\end{center}
\end {table}

\begin {table}
\caption {{\it ft} value for the $\beta^+$-decay from $^{12}N_{gs}$ to $^{12}C_{gs}$ }
\label{t:fpb}
\begin{center}

\begin{tabular}{lc} 
  & $ft$(s)
  \\ \hline 

SM(HF wf) $(0+1+2)\hbar \omega$   & 17008 \\
SM(HO wf) $(0+1+2)\hbar \omega$  &  16425\\
RPA &  3032 \\
EXP  & $13182 \pm 1.$ \cite{Ajz} \\ 
\end{tabular}  
\label{tab:4}
\end{center}
\end {table} 

\begin {table}
\caption {Same as table 1 for the exclusive muon capture rates $\Lambda_c$.}
\label{t:fpb}
\begin{center}

\begin{tabular}{lc} 
  & $\mu^{-}(^{12}C,^{12}B_{gs}){\nu}_{\mu}$ \\ 
  & $(10^{4}~s^{-1}$) \\ \hline 

SM(HO wf) $(0+1+2)\hbar \omega$  & 0.50  \\
SM(HF wf) $(0+1+2)\hbar \omega$  & 0.48  \\
SM(WS wf) $(0+1+2)\hbar \omega$ \cite{Hayes} & 0.66 \\
RPA & 2.54 \\ CRPA \cite{KL99} & 2.37(2.43) \\ 
EXP & $0.62 \pm 0.03$ \cite{Mil} \\ 
\end{tabular}
\label{tab:5}
\end{center}
\end {table} 

\begin {table}
\caption {Same as table 1 for the inclusive muon capture rates $\Lambda_c$.}
\label{t:fpb}
\begin{center}

\begin{tabular}{lc} 
  & $\mu^{-}(^{12}C,^{12}B){\nu}_{\mu}$ \\ 
  & $(10^{4}~s^{-1}$) \\ \hline 

SM(HF wf) $(0+1+2+3)\hbar \omega$ & 3.32 \\
SM(WS wf) $(0+1+2+3)\hbar \omega$ \cite{Hayes} & $(4.06)$ \\
RPA  & 5.12 \\
CRPA \cite{KL99} & 5.79(5.76) \\
EXP  & $3.8 \pm 0.1$ \cite{Suz} \\ 
\end{tabular}  
\label{tab:6}
\end{center}
\end {table}

\begin {table} 
\caption {Contribution of the most important multipolarities $J^{\pi}$ to
the inclusive $(\nu_{\mu},\mu^-)DIF$ cross sections 
$<\sigma>_{f}(10^{-40}~cm^2)$,
within the RPA, QRPA and SM approaches. 
For the latter, results obtained with two different choices of the
wavefunctions, Hartree-Fock (HF wf) and Harmonic Oscillator (HO wf) 
are shown.} 
\label{t:fpb}
\begin{center}

\begin{tabular}{lcccc} 
$J^{\pi}$ &RPA & QRPA  & SM & SM  \\
  &  & & HF wf & HO wf \\ \hline 

$0^{-}$  & 0.96 & 0.95 & 0.82 & 1.35 \\
$1^{+}$  & 4.42 & 4.48 & 2.47 & 4.11 \\
$1^{-}$  & 3.53 & 3.48 & 3.11 & 4.06 \\
$2^{+}$  & 2.04 & 1.94 & 1.38 & 2.29 \\
$2^{-}$  & 3.78 & 4.21 & 3.87 & 4.53 \\
$3^{+}$  & 1.71 & 2.25 & 1.58 & 2.33 \\
$3^{-}$  & 0.70 & 0.80 & 0.47 & 0.58 \\
$4^{-}$  & 1.36 & 1.35 & 1.11 & 1.19 \\ \\
\end{tabular}  
\label{tab:7}
\end{center}
\end {table} 

\begin {table} 
\caption {Contribution of the most important multipolarities $J^{\pi}$ to
the inclusive $(\nu_{e},e^-)DAR$ cross sections $<\sigma>_{f}(10^{-42}~cm^2)$,
within the RPA, QRPA and SM approaches. 
For the latter, results obtained with two different choices of the
wavefunctions, Hartree-Fock (HF wf) and Harmonic Oscillator (HO wf) 
are shown.} 
\label{t:fpb}
\begin{center}

\begin{tabular}{lcccc} 
$J^{\pi}$ &RPA & QRPA  & SM & SM  \\
  &  & & HF wf & HO wf \\ \hline 

$0^{-}$  & 0.3 & 0.5 & 0.6 & 0.6 \\
$1^{+}$  & 50.0 & 45.6 & 9.6 &  9.8  \\
$1^{-}$  & 1.7 & 2.0 & 2.1 & 2.2 \\
$2^{+}$  & 0.1 & 0.1 & 0.1 & 0.1 \\
$2^{-}$  & 3.0 & 3.7 & 4.0 & 3.9 \\ 
$3^{+}$  & 0.0 & 0.1 & 0.0 & 0.1 \\
\end{tabular}  
\label{tab:8}
\end{center}
\end {table}

\begin {table} 
\caption {Comparison of the
the inclusive $(\nu_{\mu},\mu^-)DIF$ cross sections
$<\sigma>_{f}(10^{-40}~cm^2)$
obtained within RPA, RPA in the same model  
space as the SM and SM with HF wavefunctions.}
\label{t:fpb}
\begin{center}

\begin{tabular}{lccc} 
$J^{\pi}$ & RPA & RPA($3 \hbar \omega$) & SM (HF wf)\\ \hline 

$0^{+}$  & 0.15 & 0.14  & 0.15  \\
$0^{-}$  & 0.96 & 0.81 &  0.82 \\
$1^{+}$  & 4.42 & 3.76 &  2.47 \\
$1^{-}$  & 3.53 & 3.14 &  3.11 \\
$2^{+}$  & 2.04 & 1.78 &  1.38 \\
$2^{-}$  & 3.78 & 3.09 &  3.87 \\
$3^{+}$  & 1.71 & 1.23 &  1.58 \\
$3^{-}$  & 0.70 & 0.40 &  0.47 \\
$4^{+}$  & 0.18 & 0.07 &  0.07 \\
$4^{-}$  & 1.36 & 1.13 &  1.11\\
$5^{+}$  & 0.39 & 0.24 &  0.15 \\
$6^{+}$  & 0.01 & 0.00 &  0.00 \\ \hline
TOTAL      & 19.23 & 15.79      & 15.18       \\ \\  
\end{tabular}  
\label{tab:9}
\end{center}
\end {table} 




\begin {table}
\caption{Results for the different processes obtained with $g_{A}^{eff}=0.9g_{A}$
both within RPA and SM.}  
\label{t:fpb}
\begin{center}

\begin{tabular}{lccc} 
 & & RPA & SM (HF wf) \\ \hline 
$(\nu_{\mu},\mu^-)DIF$ $<\sigma>_{f} (10^{-40}~cm^2)$ & exclusive & 1.76 &
0.55\\
  & inclusive & 17.02 & 13.49 \\
$(\nu_{e},e^-)DAR$ $<\sigma>_{f} (10^{-42}~cm^2)$  & exclusive & 40.79 & 6.69 \\
  & inclusive & 45.63 & 13.8 \\
{\it ft} value(s)  & & 3743 & 20278 \\
$\Lambda_c$ $(10^{4}~s^{-1})$  & exclusive & 2.05   & 0.387 \\
  & inclusive & 4.11 & 2.78 \\ 
\end{tabular}  
\label{tab:10}
\end{center}
\end {table}

\begin{figure}
\begin{center}
\epsfig{file=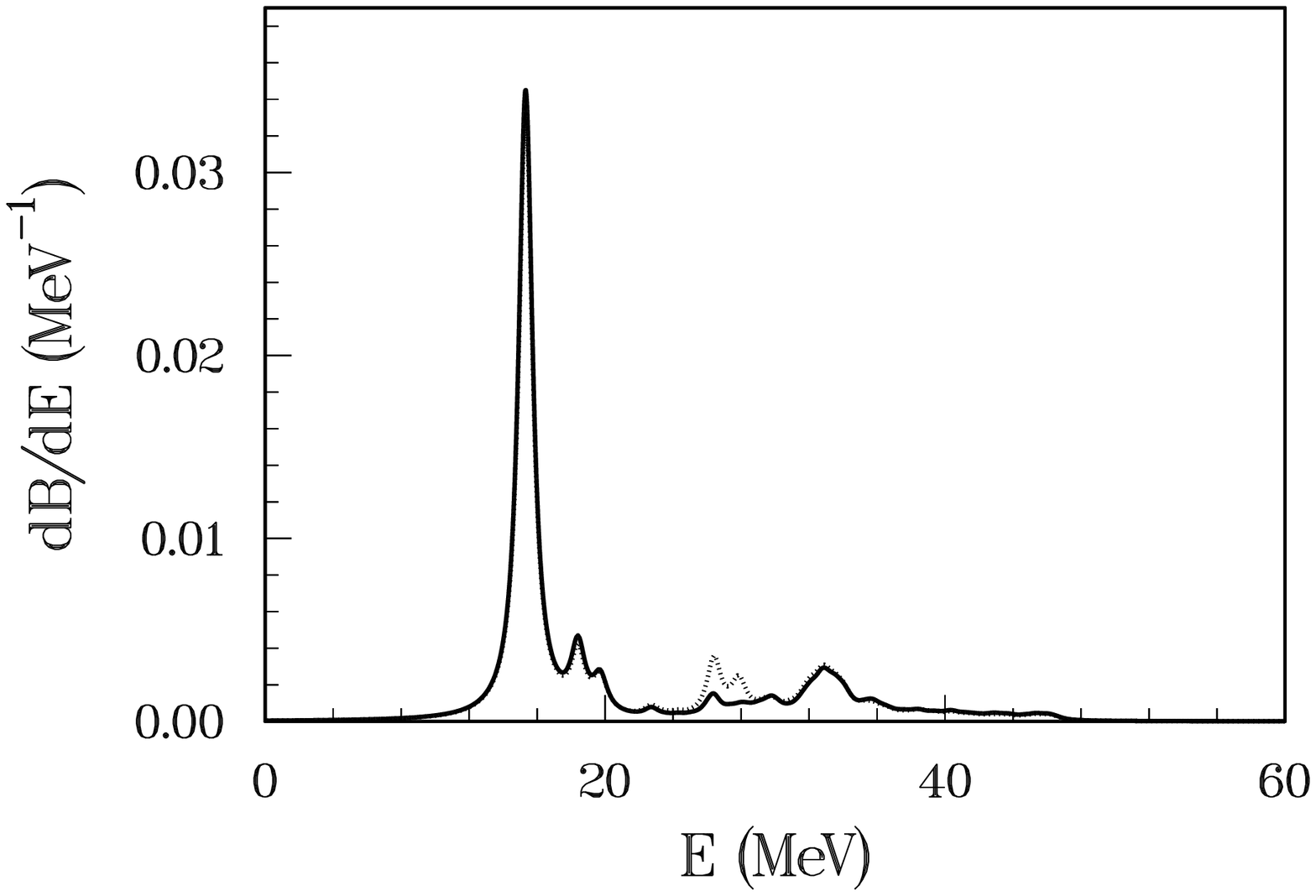,height=9.5cm}
\end{center}
\protect\caption{Strength distributions for the $J^{\pi}=1^+$ states
obtained with the transition operator
$\hat{O}=\sum_k t_+(k) j_0(q r_k)[Y_{0}(r_k) \times \sigma]^{1}$
with $q=0.2~fm^{-1}$,
using both HO (full line) and HF wavefunctions (dotted line).
The lines are the results of a folding with a lorentzian of $1~MeV$ width.}
\end{figure}

\begin{figure}
\begin{center}
\epsfig{file=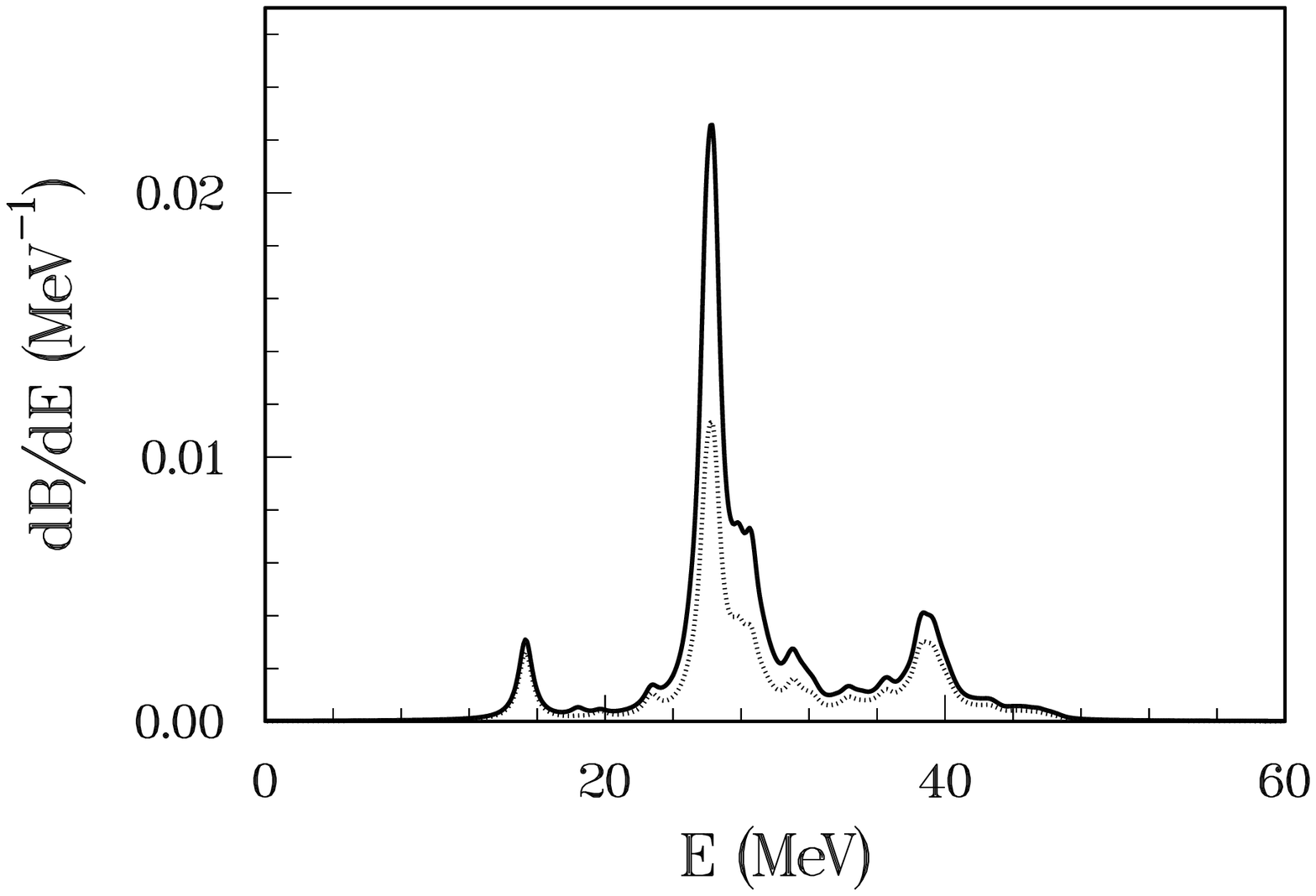,height=9.5cm}
\end{center}
\protect\caption{Strength distributions for the $J^{\pi}=1^+$ states
obtained with the transition operator
$\hat{O}=\sum_k t_+(k) j_0(q r_k)[Y_{0}(r_k) \times \sigma]^{1}$
 with $q=1.0~fm^{-1}$,
using both HO (full line) and HF wavefunctions (dotted line).
The lines are the results of a folding with a lorentzian of $1~MeV$ width.}
\end{figure}

\newpage

\begin{figure}
\begin{center}
\epsfig{file=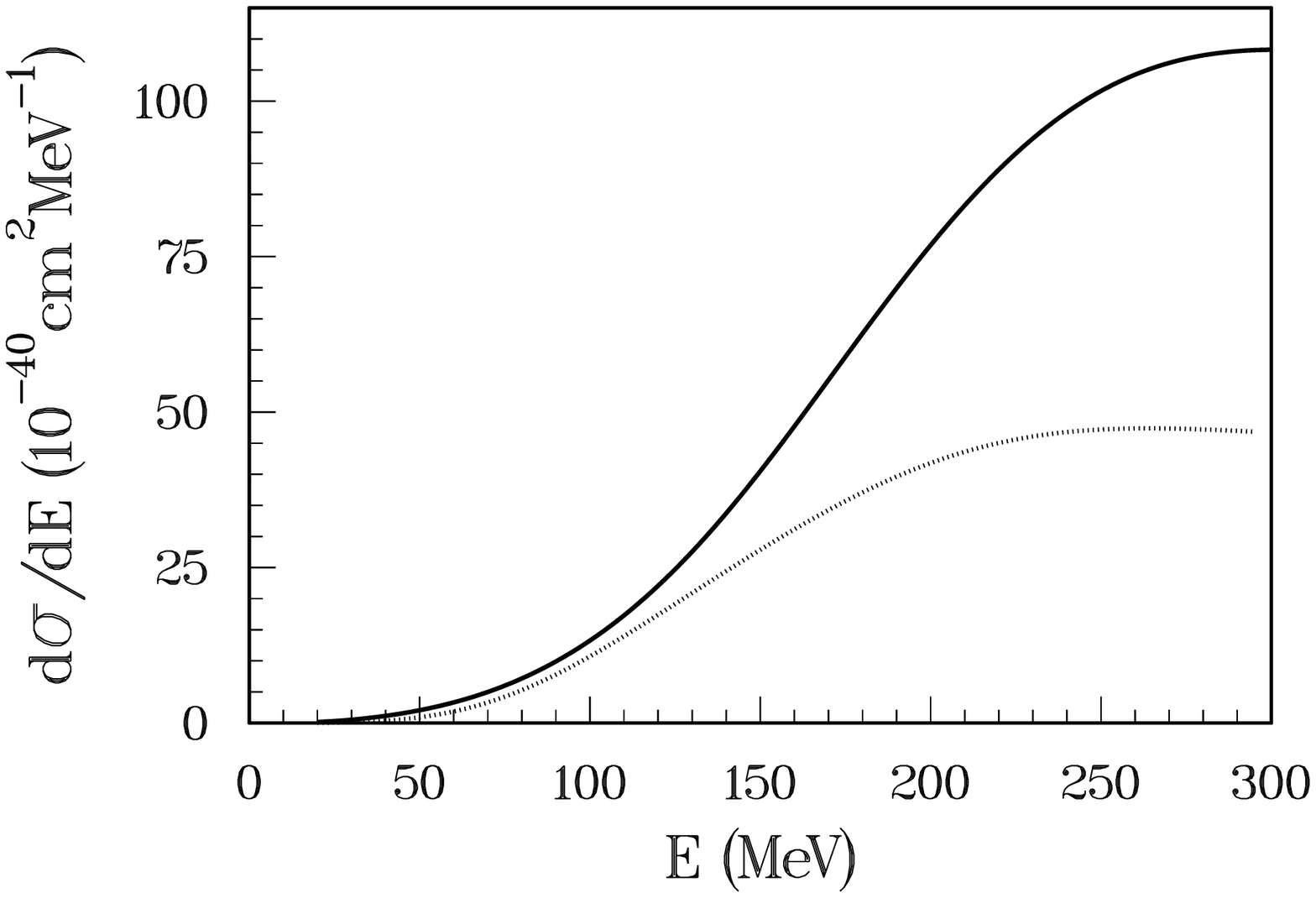,height=9.5cm}
\end{center}
\protect\caption{Inclusive $(\nu_{e},e^-)DIF$ differential cross section as a function
of neutrino energy, calculated both in RPA (full line) and
in SM (dotted line).}
\end{figure}

\begin{figure}
\begin{center}
\epsfig{file=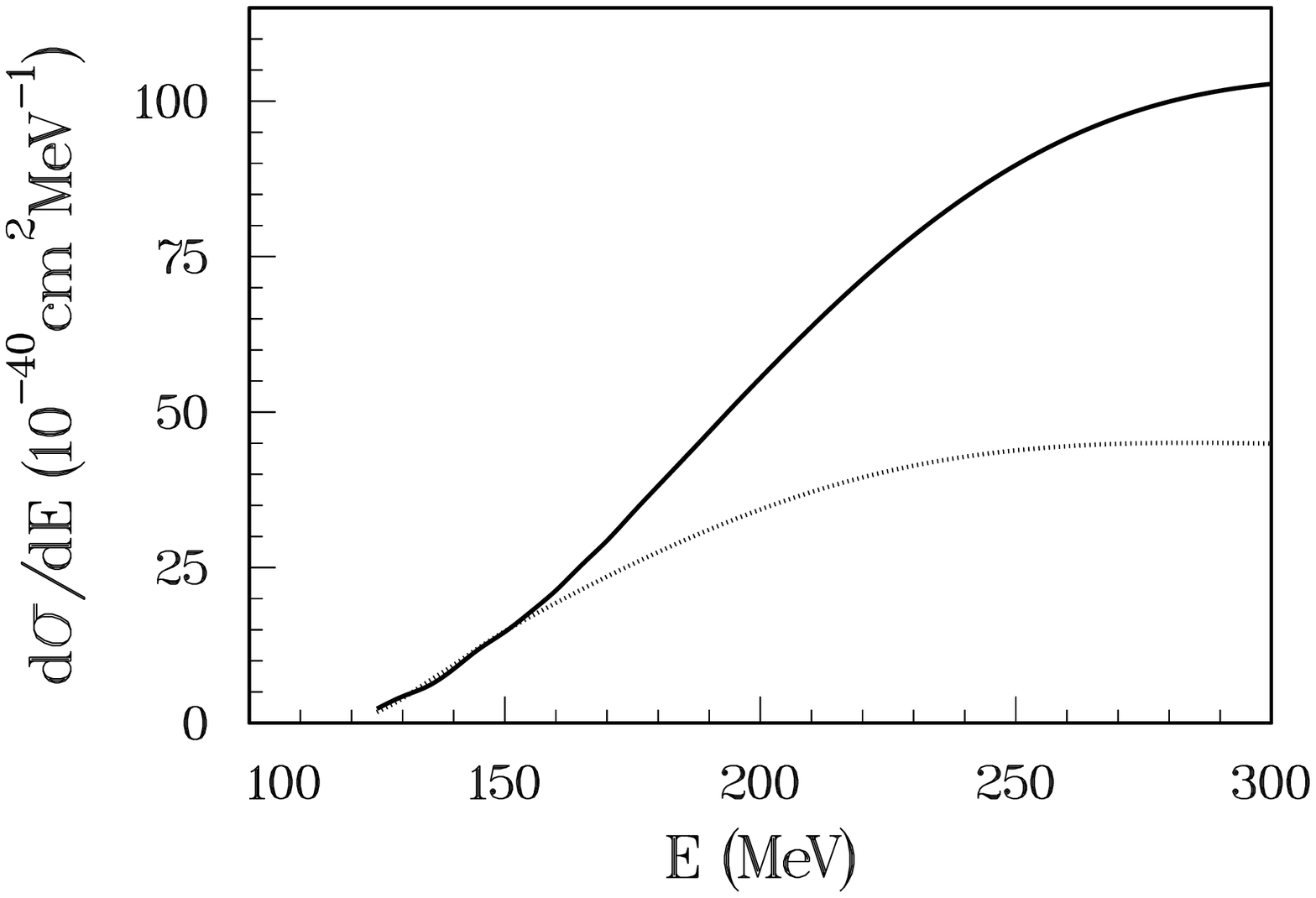,height=9.5cm}
\end{center}
\protect\caption{Inclusive $(\nu_{\mu},\mu^-)DIF$ differential cross section as a function 
of neutrino energy,  calculated both in RPA (full line) and
in SM (dotted line).}
\end{figure}

\begin{figure}
\begin{center}
\epsfig{file=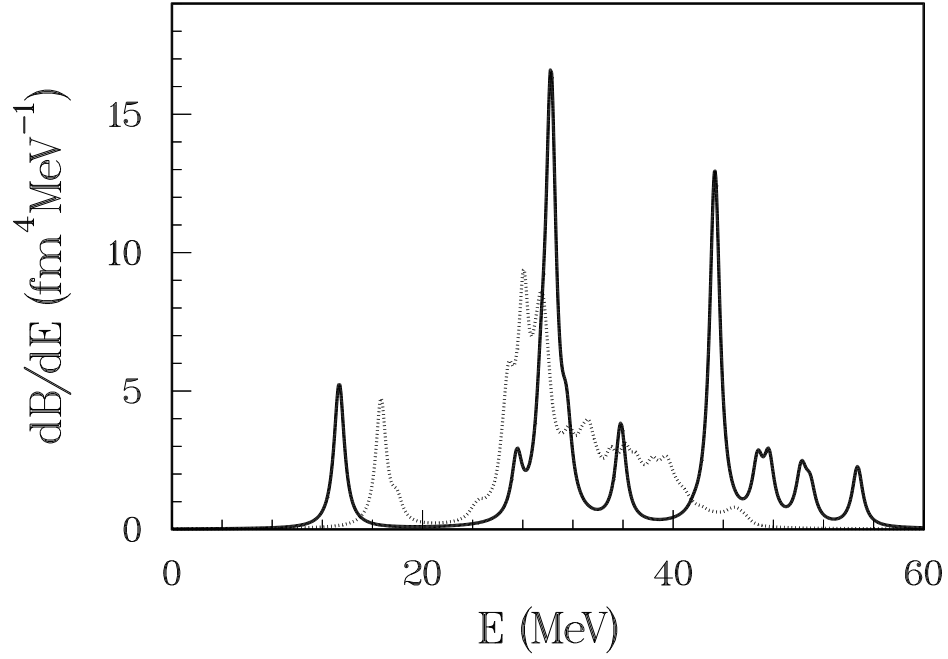,height=10.cm}
\end{center}
\protect\caption{Strength distributions 
  for the $J^{\pi}=2^+$ obtained with
the Fermi type operator $\hat{O}=\sum_k r_k^2Y_2(r_k)t_{+}(k)$ with the RPA approach (full line) and
SM (dotted line).
The lines are the results of a folding with a lorentzian of $1~MeV$ width.}
\end{figure}

\end{document}